\documentclass[twocolumn,traditabstract]{aa}

\bibpunct{(}{)}{;}{a}{}{,} %

\usepackage{graphics,hyperref,url}

\def\bea{\begin{eqnarray}}
\def\eea{\end{eqnarray}}
\def\be{\begin{equation}}
\def\ee{\end{equation}}

\def\P3{{\cal P}_t}
\def\J3{{\cal J}}
\def\T3{{\cal T}}

\def\beq{\begin{equation}}
\def\eeq{\end{equation}}
\def\bar{\begin{array}[b]}
\def\barc{\begin{array}}
\def\bart{\begin{array}[t]}
\def\ear{\end{array}}

\begin{document}
\headnote{Letter to the Editor}
\title{The neutron star in Cassiopeia A: equation of state, superfluidity, and
Joule heating} 
\titlerunning{The neutron star in Cassiopeia A: equation of state, .....}
\author{
A. Bonanno\inst{1,2}, M. Baldo\inst{2}, G. F. Burgio\inst{2}, and V.~Urpin\inst{1,3}
}
\authorrunning{A. Bonanno et al.} 
\institute{
           $^{1)}$ INAF, Osservatorio Astrofisico di Catania,
           Via S.Sofia 78, 95123 Catania, Italy \\
           $^{2)}$ INFN, Sezione di Catania, Via S.Sofia 72,
           95123 Catania, Italy \\
           $^{3)}$ A. F. Ioffe Institute of Physics and Technology,
           194021 St.Petersburg, Russia \\ 
}
\date{\today}

\abstract{
The thermomagnetic evolution of the young neutron star in Cassiopea A 
is studied by considering fast neutrino emission processes. In particular, 
we consider neutron star models obtained from the equation of state 
computed in the framework of the Brueckner-Bethe-Goldstone many-body 
theory and variational methods, and models obtained with the 
Akmal-Pandharipande-Ravenhall equation of state. It is shown that it 
is possible to explain a fast cooling regime as the one observed in 
the neutron star in Cassiopea A if the Joule heating produced by 
dissipation of the small-scale magnetic field in the crust is taken 
into account. We thus argue that it is difficult to put severe 
constraints on the superfluid gap if the Joule heating is considered.  
}

\keywords{dense matter - equation of state - stars: neutron - stars:
magnetic field - supernovae: individual: Cassiopeia A}

\maketitle

\section{Introduction}
The neutron star identified with the historical supernova SN 1680 in 
Cassiopea A (Cas A) is the youngest one to be discovered in the Milky Way.
Its estimated age of 330 yrs is in agreement with its kinematic age 
\citep{fesen06} and its thermal X-ray spectrum is consistent with 
a nonmagnetized carbon atmosphere with a surface temperature of 
$2 \times 10^6$K  \citep{ho09}. The estimated values of radius and 
mass vary from $R=8.3$ km and $M= 2.01 M_{\odot}$ to  $R=10.3$ km and 
$M=1.65 M_{\odot}$ depending on the reduction procedure of {\it Chandra 
X-ray Observatory}  archival data \citep{heinke10}. In particular, 
the authors argued that the surface temperature has rapidly decreased 
from $2.12 \times 10^6$ to $2.04 \times 10^6$ in almost one decade, 
from 2000 to  2009. This cooling rate is significantly 
stronger than expected when compared to the standard cooling model 
\citep{page06,yakovlev04} or to the medium modified Urca model 
\citep{grigorian05}. It is also unlikely that the observed cooling 
can be attributed to fast cooling models alone, such as direct Urca (DU) 
processes or neutrino emission from Bose condensates, for instance. 

In a recent work \cite{page11} interpreted the Cas A data within the 
frame of the minimal cooling paradigm \citep{page04} and explained 
that the onset of enhanced neutrino emission resulting from the 
neutron $^3P_2$ pairing in the core is enough to explain the cooling 
data. In particular, a critical temperature $T_c = 5 \times 10^8$ K for 
the triplet neutron superfluidity is implied in this process. A similar 
scenario has also been put forward by \cite{shternin11}. While in these 
works the basic idea is that neutrons have recently become superfluid 
in the core, in \cite{bla12} it has been argued that this cooling is 
produced by a reduction of thermal conductivity caused by a nuclear 
medium effect.

In this paper we argue that the data can instead be explained by dissipation
of the magnetic field and subsequent Joule heating in the surface layers
of the crust, even in the framework of fast emission processes.  
There is no evidence for the presence of a significant magnetic field 
in this neutron star. A strong field should lead to hot spots at the 
surface and spin-induced variations in the X-ray flux, which are not observed.  
However, if the magnetic field is distributed mostly on small-scales, the 
global field can be rather weak and the corresponding  variations in the 
X-ray flux might not be observable.  The existence a class of neutron stars 
with a predominant  small-scale field at the surface has been theoretically 
predicted by \cite{bonanno05,bonanno06} and subsequently observed by 
\cite{gott07} who introduced the term antimagnetars for this class of 
objects.

Rapid motions caused by hydrodynamic instabilities in protoneutron 
stars \citep{epstein79,burrows86} can lead to the onset of a turbulent 
dynamo action that amplifies the magnetic field during the first 
$30-40$ s of a neutron star life \citep{thompson93,bonanno03}.
However, if the initial rotational period is too slow this mechanism might 
not be sufficient to produce a large-scale field. On the contrary, 
turbulence in a highly conducting plasma is always accompanied by 
the generation of small-scale fields that are approximately in equipartition 
with the turbulent motions \citep{urpin04}. These fields will then be 
frozen into the crust which starts forming  soon after the end of the 
unstable stage. Therefore, the magnetic field of neutron stars can have 
a very complex geometry. In the crust, their subsequent evolution is 
determined essentially by ohmic dissipation, a relatively slow process 
because of the high conductivity of the crust. Small-scale fields can generally 
be stronger than large-scale fields and for this reason can produce more 
efficient heating in certain conditions. Estimates of a small-scale field 
are uncertain and range from $\sim 10^{15} - 10^{16}$ G to
$\sim (1-3) \times 10^{13}$ G  \citep{bonanno06,urpin04}.

\section{Joule heating in the crust}
The effect of Joule heating on the thermal evolution of neutron star was 
considered for the first time by \cite{miralles98} who argued that 
this heating can have a significantly impact on the thermal evolution of old 
neutron stars with age $t > 10^6 - 10^7$ yrs. The general relativistic 
treatment proposed by \cite{page00} led to  similar conclusions. More 
recent works \citep{urpin08,pons09} have further clarified that Joule 
heating can also be important  in very young neutron stars (with age 
$\lesssim 10^4$), thus it is important to include this effect when discussing 
the thermal evolution of Cas A. The induction equation in the crust reads
\begin{equation}
\frac{\partial \vec{B}}{\partial t} =  - \frac{c^2}{4 \pi} \nabla
\times \left( \frac{1}{\sigma} \nabla \times \vec{B} \right),
\label{uno}
\end{equation}
where $\sigma$ is the conductivity; we have neglected the Hall 
currents because these are important only for very strong magnetic 
fields $\gtrsim 10^{14}$ G.

The Joule heating term is an additional source of heat in the thermal 
balance equation which depends on the strength and configuration of the 
magnetic field. For the sake of simplicity, we assume that the radial 
length-scale of a small-scale field $L_r$ is smaller than those in the 
azimuthal and polar directions $L_{\parallel}$. 
This is reasonable because the thickness of the crust ($1-1.5$ km) 
is much smaller than the radius ($10-15$ km). This assumption does 
not qualitatively influence our results, but it simplifies calculations 
because we can neglect terms on the order of $1/r$ compared to 
$\partial / \partial r$, being $r$ the radial spherical coordinate.
It is convenient to introduce the vector potential $\vec{B} = \nabla 
\times \vec{A}$. The potential $\vec{A}$ and electric current $\vec{j}$ 
can be expanded in terms of the vector spherical harmonics 
$\vec{Y}^{(\lambda)}_{lm}$ where $\lambda = 0, \pm 1$, and $l$ and $m$ 
are the polar and azimuthal wavenumbers \citep{ak65}: 
$\vec{A} =  \sum_{\lambda, l. m} 
S^{(\lambda)}_{l, m}(r, t) \vec{Y}^{(\lambda)}_{l,m}$ and $\vec{j} = 
\sum_{\lambda, l. m} J^{(\lambda)}_{l, m}(r, t) \vec{Y}^{(\lambda)}_{l,m}$.
Then, Eq.\ref{uno} yields
${\partial S^{(\lambda)}_{jm}}/{\partial t} = \frac{c^2}{4 \pi \sigma} 
{\partial^2 S^{(\lambda)}_{jm}}/{\partial r^2}$ \citep{urpin94}. 
Continuity of the field at the stellar surface $r=R$ leads 
to the 
boundary condition $\partial S^{(\lambda)}_{jm} / \partial r \approx 0$
with the 
accuracy in terms of the lowest order in $L_r / L_{\parallel}$. 
The spherical components of $\vec{j}$ can be expressed in terms of 
$S^{(\lambda)}_{l, m}$. 
Then, the rate of Joule heating is $\dot{q} = j^2/\sigma$ and in general 
it will depend on $\theta$ and $\varphi$, the polar and azimuthal 
spherical coordinates. The cooling codes for neutron stars usually do
not take into account departure from spherical symmetry and we will thus 
use the angle-averaged expression for $\dot{q}$. Using orthogonality of 
the vector spherical harmonics, we finally obtain 
\vspace{-0.3cm}
\begin{equation}
\dot{q} = \frac{c^2}{16 \pi^2 r^2 \sigma} \sum_{\lambda, l. m} \left(\frac{
\partial^2 S^{(\lambda)}_{l, m}}{\partial r^2} \right)^2 .
\label{due}
\end{equation}
{In our model, $\dot{q}$ depends on the radial distribution of the 
small-scale field. At the beginning of the evolution, the instabilities
occur almost in the whole volume of a neutron star. Hence, motions 
caused by instabilities generate small-scale fields almost everywhere 
within the neutron star. The characteristic length-scales of these 
fields range from the main turbulent scale ($\sim 1$ km) up the 
dissipative length scale. During the cooling these fields will be frozen 
into the crustal matter and their evolution is determined mainly by 
ohmic dissipation. Dissipation in the core is very slow because of 
large $\sigma$, and we neglect it. In the crust, fields with different 
length-scales $\lambda$ located at different depths decay on different 
time scales, $\tau_{\lambda}\sim 4 \pi \sigma \lambda^2 / c^2$. 
Depending on the age, the main contribution to $\dot{q}$ is given by 
fields with different characteristic length scales. Using the expression
for $\tau_{\lambda}$, it is easy to estimate that the main contribution 
to $\dot{q}$ at the age $\sim 300$ yrs is determined by magnetic fields 
with a characteristic length scale $300-600$ m. Fields with smaller 
characteristic length scales have already disappeared, but those with 
greater length scales do not contribute yet to the Joule heating. Therefore, 
in the sum in Eq.(\ref{due}) we can consider only a dominant term with 
the characteristic radial length scale $300-600$ m and the 
corresponding field strength.}

\section{Equation of state}
In this paper, we concentrate on the thermal evolution of a neutron star 
with the fast neutrino emission processes. 
We consider two models based on different equation of state (EOS) to be 
representative of these stars.
The first is derived within the Brueckner-Bethe-Goldstone many-body 
theory \citep{baldo99}, the BHF EOS. In this many-body approach one 
introduces the Brueckner reaction matrix G, which can be interpreted as 
the effective in-medium nucleon-nucleon (NN) interaction, and the single 
particle potential $U(k;\rho) = \sum _{k'\leq k_F} \langle k k'|G(\rho)|k k'
\rangle_a $ . The subscript ``{\it a}'' indicates antisymmetrization and 
$\rho$ is the baryon density. From the potential $U$ the EOS can be 
obtained, and the in-medium nucleon effective mass $m^*$ at the Fermi 
surface can be derived according to the relation
${m^*}/{m} =1/( 1 \,+\, \frac{m}{\hbar^2 k_F} \frac{d U}{d k} )$
where the derivative of $U$ is taken at $k = k_F$.
The accuracy and convergence of the expansion have been extensively 
studied \citep{song98,baldo00} and it turns out that in order to 
reproduce the correct saturation point of symmetric nuclear matter one 
needs to introduce three-body force, which are reduced to  density 
dependent two-body force by averaging over the position of the third 
particle \citep{baldo97}. We took the Argonne $v_{18}$ as NN potential 
\cite{wiringa95}, supplemented by the so-called Urbana model 
\citep{carlson83} as the three-body force. The saturation point
($\rho_0 \approx 0.16~\mathrm{fm}^{-3}$, $E/A \approx -16$ MeV) is then
well reproduced, and incompressibility and symmetry energy are compatible
with those extracted from phenomenology \citep{taranto13}. The second 
EOS is the so-called APR EOS by \cite{apr} obtained from the Argonne 
$v_{18}$ NN interaction and a variational procedure as well as Urbana 
three-nucleon interactions. Although the BHF and variational methods 
are connected \citep{baldo12}, they give different EOS at high density. 
We assume that the matter is composed of neutrons, protons, and leptons 
in beta-equilibrium where electrons and muons are treated as relativistic 
Fermi gas.

The two EOS start to deviate at nucleon densities $> 2 \rho_0$ and the 
symmetry energy for APR EOS becomes smaller than for the BHF EOS. The 
symmetry energy determines the proton fraction, hence the onset of the 
DU process. In the BHF approach it already takes place at 3$\rho_0$, 
whereas in the APR case it is allowed at larger density, close to 
1$\rm \; fm^{-3}$ \citep{zhou04}. Therefore, in the BHF case NS with mass 
around 1.2$M_\odot$ can cool rapidly through DU, whereas  in the APR 
case DU is allowed only for heavy NS, close to the maximum mass. We note 
that the values of the maximum allowed mass is different in the two 
approaches, being slightly above 2$M_\odot$ in the 
BHF case and close to 2.2 $M_\odot$ in the APR case \citep{taranto13}.
\begin{figure}
\begin{center}
\includegraphics[width=7cm,height=5.5cm]{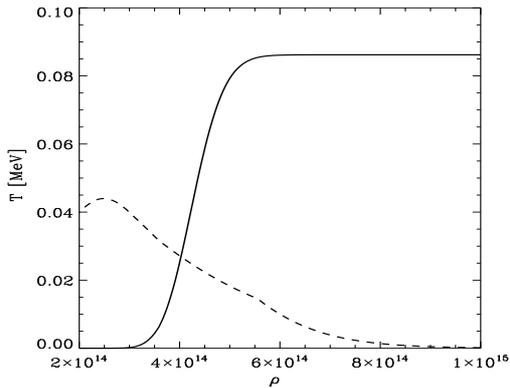}
\caption{Critical temperatures of the neutron $^3P_2$ gap (solid line) as 
a function of the density and of the proton $^1S_0$ gap (dashed line) 
used in the computation for the $1.4 \;M_{\odot}$ neutron star.}
\end{center}
\end{figure}
%
\begin{figure*}[tb]
\begin{center}
\includegraphics[width=17cm,height=6.5cm]{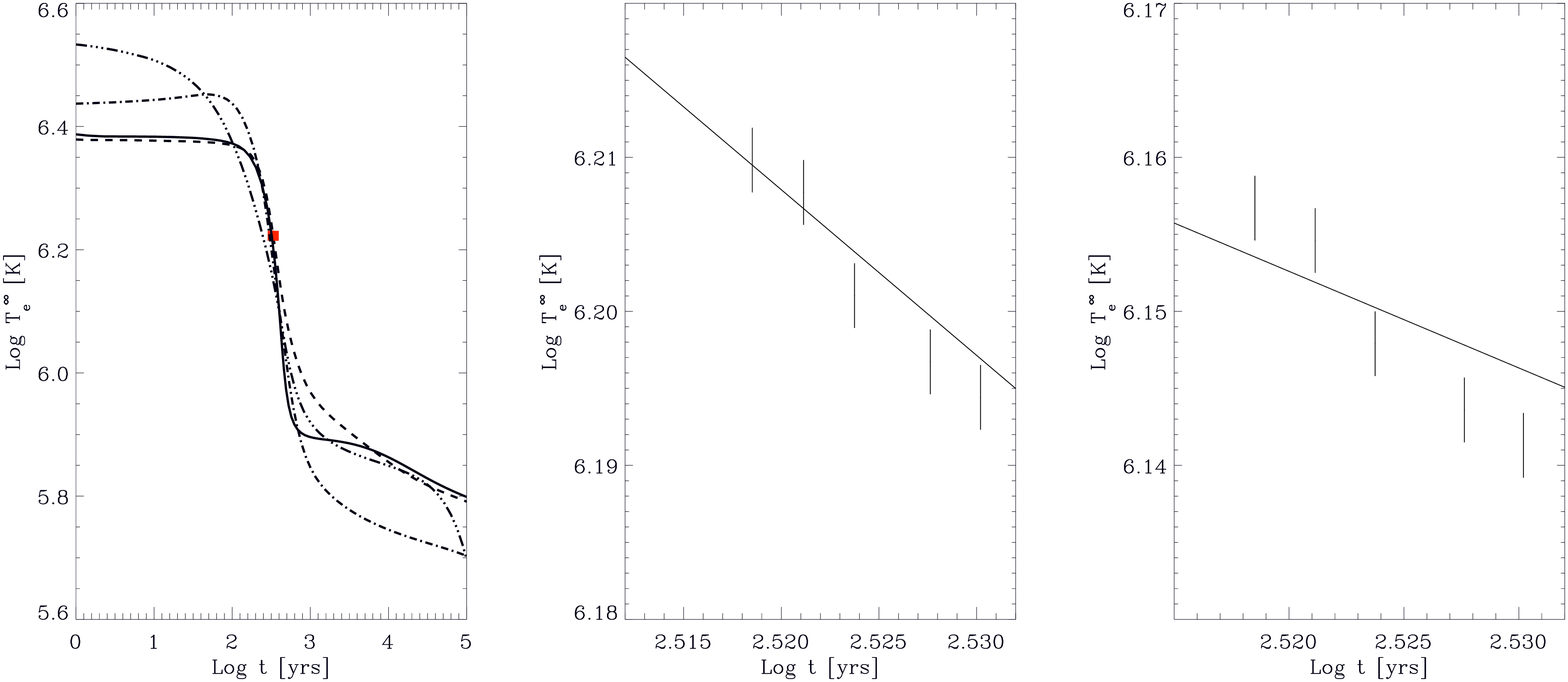}
\caption{Left panel: Redshifted effective temperature $T_e^{\infty}$ vs time 
for the neutron star models with different EOS and masses. The solid, 
dashed, and dot-dashed curves correspond to the neutron stars of 1.4, 
1.6, and 2.0 $M_{\odot}$, respectively, 
with the  BHF EOS. The dot-dot-dashed line represents 
2$M_{\odot}$ neutron star with the EOS of APR.
Central panel: fit of the Cas A data with the BHF EOS 
for a neutron star of $1.4 \;M_{\odot}$ and a turbulent magnetic 
field of initial strength $2.2 \times 10^{13} $G. 
Right panel: fit of the Cas A data with the APR EOS 
for a neutron star of $2.0 \; M_{\odot}$ 
and a turbulent magnetic 
field of initial strength $9.5 \times10^{12} $G. 
}
\end{center}
\end{figure*}

\section{Numerical results}
Our numerical simulations use the public code {\em NSCool}\footnote{\url{http://www.astroscu.unam.mx/neutrones/NSCool/}}
in which  the magnetothermal evolution has been taken into account by coupling 
the induction equation with the thermal equation via the Joule heating 
source. In particular, the induction equation was 
solved via an implicit scheme and the electron conductivity was
calculated with the approach described in \cite{potekhin99}. We assumed 
that the composition of the crust corresponds to the so-called accreted 
matter with an envelope mass of $\Delta M=3.6 \times 10^{-15} M_\odot$  
and an impurity parameter $Q=0.01$. 

The critical temperature of $^3 P_2$ pairing of neutrons is $10^9$ K 
in all our models. In particular, we followed the estimate of \cite{baldo98}
for the proton and neutron superfluidity where the gap extends up to 
the center of the neutron star. For actual computations we have thus 
modeled the density dependence of this gap with  a sharp increase after 
$\rho \sim 4 \times 10^{14}$ gm/cm$^3$ which extends up to the center of 
the star in all the models (see Fig.1 for the $1.4 \;M_{\odot}$ example). 
 
{The small-scale magnetic field is assumed to be anchored in the crust and 
its characteristic length scale should vary around 
$\lambda_r \sim 300-600$ m as estimated in the previous section. We expect 
that the initial field strength lies in the range $ (1-3) 10^{13}$ G, 
according to the estimate of \cite{bonanno06} for a dynamo generated 
small-scale field. The crucial feature, from the observational point of 
view, is the occurrence of a sharp transition, a transit time
\citep{page11} of a few decades, during which the cooling can be characterized 
by an almost constant  slope $s = - d \log T^\infty_e/d \log t$, as shown 
in the center and right panels in Fig.~2. 
In determining the observed slope and effective temperature, the strength 
of the magnetic field and its length scale play antagonistic roles. By 
increasing the strength, it is possible to slow down the cooling, and the 
slope $s$ decreases. On the contrary, by increasing the length scale it 
is possible to decrease the efficiency of the Joule heating. This is 
expected since $\dot{q}\propto (\partial S/\partial r)^2$, 
and therefore the greater the gradients of $S$ are, the more efficient  
the Joule heating will be. In our numerical simulations we tried to find 
the best compromise between the field strength and its length scale 
in order to explain both the slope and the observed effective temperature
of Cas A. 

Our results are summarized in the three panels in Fig.~2, where our best 
models with different EOS are displayed. We found that the length scale 
of the initial magnetic field should be $\sim 600$ m in the models 
with the BHF EOS and $\sim 300$ m in the model with the APR EOS. It is 
reassuring to notice that these values are comparable to the main length scale 
of turbulence in protoneutron stars, $\sim 1$ km ( see, e.g., 
\cite{burrows86}).

On the other hand, the obtained values of the field are 2.2, 1.3, and 
$1.5 \times 10^{13}$ G for models with the BHF EOS and with 1.4, 1.6, and 
2.0 $M_{\odot}$, respectively, while  for the model with the APR EOS, we 
obtain $9.5 \times 10^{12}$ G, in agreement with the estimate discussed 
in \cite{bonanno06}.}

\section{Discussion}
Most probably all neutron stars have small-scale magnetic fields amplified 
in the course of a short unstable stage soon after collapse. Turbulent motions 
generate small-scale fields on a very short time scale $1-10$ ms. 
The strength of these fields is $ (1-3) \times 10^{13}$ G at the end of a 
convective stage. Generation of a global magnetic field is determined by 
rotation and can be suppressed if the neutron star rotates slowly at birth. 
Therefore, there should exist a particular class of neutron stars that 
has relatively strong small-scale fields and has no (or a very weak) global 
field. Perhaps, Cas A is representative of this class. 

{Dissipation of small-scale fields $\sim 10^{13}$ G can lead to 
departures from the standard cooling scenario of non-magnetic neutron stars. 
Our calculations show that Joule heating can provide a sufficient amount of
heat to account for the temperature and cooling rate of Cas A even for the 
neutron star models with fast cooling. Since the presence of small-scale 
fields is a general feature of neutron stars, our model predicts that
departures from the standard cooling should be detected in all young
stars. According to our calculations, the cooling rate of Cas A will 
remain approximately unchanged during at least a few hundred years. 

It is important to stress  that in our calculation the presence of 
a gap extending up to the center was needed only to suppress the fast 
cooling due to the DU, otherwise it did not play an essential role in 
explaining the rapid cooling during the transit phase. 
In this model it is impossible to  put severe constraints on $T_c$ from the observed slope  during this phase. In particular, although in actual calculations the 
critical temperature $T_c = 10^9$ K for the triplet neutron superfluidity 
has been used, we found that a successful fit can also be obtained with 
$T_c = 5 \times 10^8$K. Generally, a good fit for the temperature and 
cooling rate can be obtained for other values of $T_c$ by varying 
the magnetic field and its length scale.
}

While this work was in progress a new analysis of the Cas A appeared
\citep{els13} where the cooling of Cas A is less extreme than previously 
reported. Clearly, all our conclusions would remain valid even in this case, 
provided the strength of the initial magnetic field is slightly reduced.

{\it Acknowledgments. 
VU thanks INFN-Sezione di Catania and INAF-Osservatorio Astrofisico di 
Catania for hospitality and financial support. AB would  like to thank Dany Page for his help in using 
the public code {\it NSCool}}


\end{document}